# Mega-MUSE Nearby Galaxy Serendipity Survey

White Paper in response to ESO call for ideas

"What science questions will astronomy need to answer in the 2040s?"

Martin M. Roth [1,2,3]

Affiliations:

(1) Leibniz-Institut für Astrophysik Potsdam (AIP), An der Sternwarte 16, 14482 Potsdam, Germany
(2) Institut für Physik und Astronomie, Universität Potsdam, Karl-Liebknecht-Str. 24/25, 14476 Potsdam, Germany
(3) Deutsches Zentrum für Astrophysik (DZA), Postplatz 1, 02826 Görlitz, Germany

Date of submission: December 14, 2025

## Context

The idea put forward in this White Paper is the notion that novel observing facilities for astronomy almost always make discoveries that were not anticipated at the time of planning and conceptual design. A prominent more recent example of this kind is JWST's unexpected discovery of disk galaxies at z>3, (e.g., Kartaltepe+2023; Ferreira+2022, 2023; Robertson+2023; Nelson+2023; Jacobs+2023), or the discovery of the massive grand design spiral *Alaknanda* at z≈4 by Jain & Wadadekar (2025): a paradigm change since the observation of high redshift galaxies with HST.

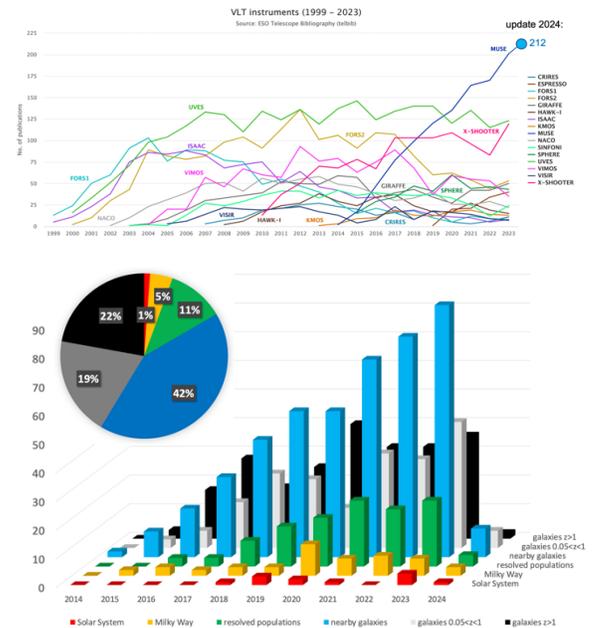

Fig.1: VLT instrument productivity [papers per year] (top), MUSE publications by targets: resolved stellar pop. in green (bottom).

MUSE as an innovative optical integral field spectrograph at the ESO VLT was commissioned in 2014 and became soon the most productive instrument at Paranal (Roth 2024a), Fig. 1 (top). It has become **transformative** in the sense that beyond its primary objective of deep integral field spectroscopy (IFS) for extragalactic work and cosmology, the invention of **crowded field PSF-fitting IFS** (Kamann+2013) has revolutionized our ability to obtain high quality spectra of severely crowded fields in the nearby universe – analogous to the revolutionary impact of CCD crowded field photometry in the 1980s (Stetson 1987), that has become instrumental for the study of stellar evolution in globular clusters (GC). Fig. 1 (bottom) from Roth (2024a) illustrates how over 10 years since the advent of MUSE the spectroscopic analysis of resolved stellar populations has become an emerging new field with steady growth (green symbols).

MUSE observations for specific goals, e.g., binary fraction in GCs (Giesers+2018,2019), massive stars in low metallicity dwarf galaxies (Evans+2019), extragalactic planetary nebulae (Soemitro+2023), etc., have yielded serendipitous discoveries of rare stars, ionized nebulae, and other objects that have escaped the detection from previous broad- and narrow-band imaging surveys. This White Paper makes the case for the exploitation of a future **Mega-MUSE facility at a 10m+ aperture telescope supported by ground layer adaptive optics** for the systematic search for rare objects in nearby galaxies whose statistics are uncertain due to selection effects. The discovery potential of such a facility is **unprecedented** and **transformative**, yet by its very nature unpredictable.

## Scientific Objective

An IFS nearby galaxy serendipity survey is to be distinguished from any other survey program in that it is not targeting the solution of a single specific astrophysical problem, but enables the exploitation of data obtained from other surveys in **Local Group (LG) and Local Volume (LV) galaxies** to clarify open science questions that **cannot be solved otherwise**. The most prominent objective is to obtain **solid statistics of rare stars and gaseous nebulae** as well as their properties that are not accessible in the Milky Way disk due to extinction and selection effects and **will not become available over the next 15 years**. Recent experience with archival data from the unique large field-of-view (FoV) integral field spectrograph MUSE at the VLT has already demonstrated the scientific value of **datasets that were initially obtained for another purpose**. The ESO archive has already yielded the discovery of countless H II regions, supernova remnants (SNR), planetary nebulae (PN), luminous stars of spectral types from O to M down to luminosity class III, carbon stars, WR stars, Be stars, symbiotic stars, microquasars, etc. in galaxies of the LG, or further out into the LV. Any deep exposure with IFS in nearby galaxies will capture serendipitously numerous extended objects and point sources, both in the continuum and in emission. As opposed to direct imaging, even with narrow-band filters, IFS reaches unprecedented sensitivity for such objects due to its high spectral resolution and broad free spectral range. However, even the comparatively large $1 \times 1$ arcmin$^2$ FoV of MUSE is too small to cover the footprint of LG or nearby LV galaxies within reasonable observing times. A new survey IFS facility providing an order of magnitude larger FoV would change this situation. Given the productivity and the current oversubscription of MUSE it is not entirely unrealistic to assume the implementation of a Mega-MUSE survey telescope based on a concept like the one from Pasquini+2016. **A coherent approach to systematically analyzing data cubes** from such a facility to create source catalogues and identify unusual objects will be of unique value for the **statistics of rare objects**, of stars (in particular the initial mass function), nebulae, and the **discovery of unknown phenomena**. No other facility in the foreseeable future will have that transformative capability. In what follows, a few examples will illustrate the qualitative and quantitative value of such a survey.

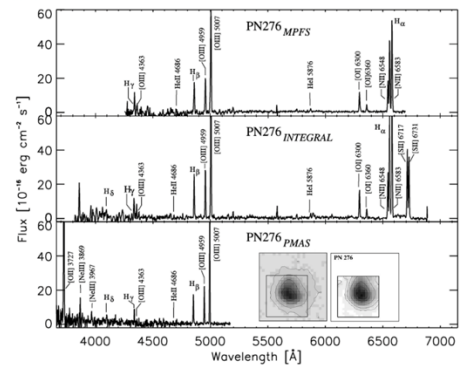

Fig. 2: Integral field spectroscopy of emission line source in M31, misclassified as a PN, but

Arguably the first serendipitous discovery of an extragalactic SNR through IFS is the object PN276 in M31 that had initially been misclassified as a planetary nebula on the basis of narrow-band imaging (Jacoby+ 1990; Roth+2004). Observations with the 1$^{st}$ generation IFUs MPFS (Sil'chenko+2000), PMAS (Roth+2005), and INTEGRAL (Arribas+1998) revealed that this object is in fact extended and shows the typical emission line spectrum of a SNR with the characteristic bright [S II] lines $\lambda\lambda$ 6717, 6731 (Fig. 2). Kreckel+2017 have made use of this property in their distance determination of NGC 628 with MUSE by measuring the planetary nebula luminosity function (PNLF) in [O III] $\lambda$ 5007, and rejecting SNR interlopers on the basis of their spectra.

Based on MUSE observations, Evans+2019, presented the first stellar spectroscopy in the low-luminosity, low metallicity dwarf galaxy Leo P. It was chosen for its low oxygen abundance (3% solar) and proximity (~1.6 Mpc) as a sample galaxy for massive stars with near-primordial compositions similar to those in the early Universe. One pointing of 6.7 hrs resulted in a total of 341 spectra, out of which 32 were classified as spectral type OB, 5 as AGB star candidates, 6 as spectral types K...M, and 298 spectra remaining unclassified. As an unexpected finding, the MUSE data also revealed two 100 pc-scale ring structures in Hα emission.

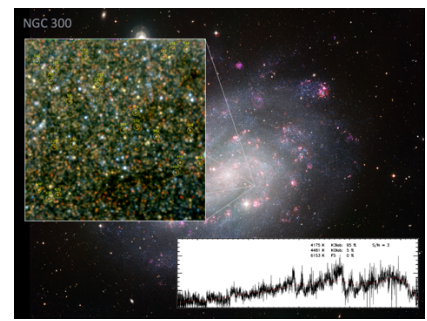

Fig. 3: MUSE observations in NGC 300.

A GTO pilot study in NGC 300 was conducted to demonstrate that IFS of stellar fields with high spatial resolution and excellent seeing conditions reaches unprecedented depth in severely crowded fields (Roth+2018). Point spread function-fitting IFS using the PampelMuse code (Kamann+ 2013) yielded deblended spectra of individually distinguishable stars, thus providing a complete inventory of blue and red supergiants, and asymptotic giant branch stars. The study yielded a catalog of luminous stars, rare stars such as WR, and other emission line stars, carbon stars, symbiotic star candidates, PNe, H II regions, SNR, giant shells, peculiar diffuse and filamentary emission line objects, also background galaxies, along with their spectra. Fig. 3 illustrates the initially unex-pected finding, that MUSE is very efficient in detecting carbon stars that can only be classified on the basis of their spectra. Using the same methodology, a carbon star was found in the ultra-faint dwarf galaxy Eridanus 2 (Zoutendijk+2020) in search for hypothetical MACHOs in the central star cluster, thus providing evidence for an intermediate age stellar population, that would not be expected in this kind of galaxy.

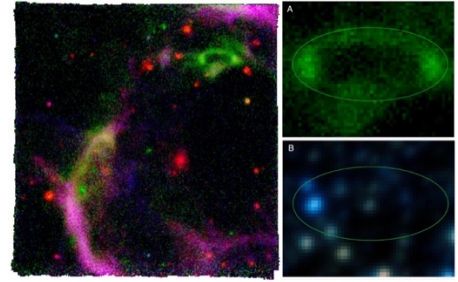

Taibi+2024a conducted a MUSE study in IC 1613, a star-forming, gas-rich, and isolated LG dwarf galaxy (d=759 kpc). The analysis yielded accurate classifications and line-of-sight velocities for about 800 stars. The sample includes not only RGB and main sequence stars, but also a number of Be and C stars. Reliable metallicities ($\delta_{[Fe/H]}$ ~ 0.25 dex) were obtained for about 300 RGB stars. The kinematic analysis of IC 1613 revealed for the first time the presence of stellar rotation with high significance, in general agreement with the rotation velocity of the neutral gas component. In the same dataset, Taibi+2024b, reported the detection of a highly ionized extended source close to the galaxy center showing bright emission in [O III] $\lambda\lambda$ 4959, 5007 Å, in $H_\alpha$, $H_\beta$, and [S II] $\lambda\lambda$ 6717 and 6731 Å (Fig. 4). The physical nature of this enigmatic object is entirely unknown.

Fig. 4: Emission-line map of the MUSE pointing in IC 1613; $H_\alpha$ red, [S II] blue, and [O III] green.

The examples above were chosen randomly for illustration and are by no means a representative or complete account of serendipitous discoveries made with MUSE in the first decade of operation. A more complete review of similar results from the first five years of operation was presented by Roth+2019. For the detection of 40.000 gaseous nebulae in the PHANGS survey, see Congiu+2023.

## Instrumental requirements and disruptive technologies

In order to achieve the required order of magnitude gain over MUSE/BlueMUSE, the envisaged Mega-MUSE should be operated at a **10m+ class survey telescope** with **ground-layer adaptive optics** support at a **stable** stationary focus. It should offer a **1000 × 1000 spaxel** integral field unit with a spaxel size of **0.25 × 0.25 arcsec²** and a wavelength range of **350…900 nm** at a resolving power of **R ≈ 5.000**. IFS should be the **exclusive mode of operation** with a fixed configuration. The instrument should allow for stability reasons **no moving parts** such as filter or grating exchange mechanisms. The instrument should deliver a similar or better performance in terms of **throughput and stability** in comparison to MUSE. MUSE/BlueMUSE are already very complex instruments. The Mega-MUSE configuration would be even more complex, therefore requiring technology development (Bacon+2023), especially for the **detector system**. Detectors should be revolutionary in comparison with the classical CCD technology (bulky vacuum/cryogenics/controllers). It should employ the currently emerging detector technology of extremely low noise CMOS sensors (Krynski+2025, Roth 2024b), SPAD, or equivalent, in order to allow for **post-facto rebinning in wavelength** without noise penalty for very faint sources.